# Optimizing Economic Markets through Monte Carlo Simulations and Magnetism-Inspired Modelling

Chee Kian Yap[1] and Arun Kumar Singh[2]
[1]Queensland University of Technology, Australia
[2]ICFAI University, Raipur, India

**ABSTRACT.** This study presents a novel approach to modeling economic agents as analogous to spin states in physics, particularly the Ising model. By associating economic activity with spin orientations (up for inactivity, down for activity), the study serves to optimize market dynamics using concepts from statistical mechanics. The aim is to use Monte Carlo simulations and maximize surplus by allowing the market to evolve freely toward equilibrium. The introduction of temperature represents the frequency of economic activities, which is crucial for optimizing consumer and producer surplus. The government's role as a temperature regulator (raising temperature to stimulate economic activity) is explored for market optimization. Results from simulations and policy interventions, such as introducing a "magnetic field," are discussed, showcasing complexities in optimizing economic systems while avoiding undue control that may destabilize the markets. The study provides insights into bridging concepts from physics and economics, paving the way for a deeper understanding of economic dynamics and policy interventions.

**1 Introduction**

In economic systems, where human agents interact as participants in market activities, parallels can be drawn with physical phenomena such as magnetism. Just as magnetic spins can be modeled as either up (inactivity) or down (buy), economic agents can be seen as engaging in economic activities and behaviors. This analogy lays the groundwork for exploring the optimization of market dynamics through thermodynamic principles, akin to how spins align or anti-align in magnetic materials.

In the Ising model of magnetism, spins interact with each other based on their orientation, influenced by the coupling constant *J*. Similarly, economic agent decisions are influenced by market conditions, policies, and individual preferences. At low temperatures (representing low economic activity), agents tend to adopt a state of minimal activity (up state), reflecting limited spending or investment due to uncertainty or economic constraints. Conversely, as temperature (economic activity) increases, agents become more active in buying activities to optimize their economic surplus. This transition from a low-activity state to a more dynamic one can be simulated using Monte Carlo methods, facilitating the study of how market optimization evolves.

One key insight to this optimization is that raising the "temperature" of economic activity allows the market to evolve freely, leading towards an equilibrium where surplus is maximized. This concept aligns with thermodynamic principles, where a system tends to move towards a state of maximum entropy (maximum surplus in economic terms) when allowed to evolve freely. Achieving this equilibrium requires careful consideration, especially in large-scale economic systems. Like raising the temperature in a thermodynamic system, government interventions can influence market dynamics significantly. Yet, such interventions must be designed to benefit the entire system rather than serving individual interests. The notion of a "ground state" in economic terms reflects the state where all agents are either inactive (up state) or fully engaged (down state), analogous to the alignment of spins in a magnetic material. Notably, both extreme states (all inactive or all active) are suboptimal in economic systems which indicates an imbalanced market condition. The ideal scenario for market optimization lies in a balanced mix where many agents are active, driving economic activity without leading to instability or inefficiencies.

Unlike spins in a material, economic agents are interconnected in complex networks, influencing and being influenced by multiple factors simultaneously. Here, the role of government becomes pivotal. Governments act as regulators and influencers, akin to providing the "temperature" for the economic system. However, their interventions must maximize overall surplus rather than serve narrow interests. Policies promoting fair competition, transparency, and financial stability create an environment where market participants can optimize their activities and contribute to economic growth.

This study aimed to bridge concepts from thermodynamics and economics, using mathematical models and simulations to understand how economic systems evolve and can be optimized. Monte Carlo simulations were used to study the effects of temperature (economic activity levels) and external interventions (government policies) on market dynamics and surplus optimization. Blending insights from physics and economics enables a deeper understanding of complex economic systems to create more efficient, equitable, and sustainable markets. Interdisciplinary approaches are used to understand complex systems, bridging concepts from physics to economics. One intriguing analogy is modeling economic agents as particles in a magnetic system, where buying actions are analogous to spin down and non-buying actions as spin up. This paper examines economic dynamics through the lens of thermodynamic principles, using Monte Carlo simulations to better identify market optimization strategies.

## 2 Methodology

**Modeling Economy Agents as Spin States**: The work suggests using a model like spin states in physics, where "down" corresponds to buying activity and "up" corresponds to no activity in the economic context. This analogy is akin to the Ising model in physics, where spins can be pointed up or down.

**Ground State and Temperature**: In the Ising model, the lowest energy state is the ground state. In this economic model, the ground state corresponds to all agents doing nothing (up). When temperature (representing market activity or energy in the system) increases, agents (spins) are more likely to participate in the market in buying.

**Monte Carlo Simulation**: This is a computational technique used to simulate complex systems by random sampling. In this context, it could be used to simulate how economic agents (spins) interact and evolve based on their buying or non-buying behaviors.

**Optimizing the Market**: The aim is to optimize the market by allowing it to evolve freely at higher temperatures. This is akin to the annealing process, increasing market activity to reach equilibrium, where there is maximum surplus.

**Reservoir Analogy**: By raising the temperature (representing market activity) and ensuring entropy increases while keeping temperature change close to zero for stability, the system is likened to a reservoir that can absorb and dissipate energy while keeping temperature constant.

$$\Delta S > 0 \qquad (1)$$
$$\Delta T = 0 \qquad (2)$$

**Magnetism in Economic System**: The analogy drawn here relates to how magnetism deals with spin states and energy levels, which can be analogous to economic activities such as buying. The Normalized magnetization in this model is described as:

$$M = \frac{|Buyers - Nonbuyers|}{Total\ Agents} \quad (3)$$

**Ising Model Challenges**: While the Ising model can represent binary states (buy or non-buying), it struggles to capture the nuances of economic behaviors, such as the effect of full connectivity of one agent to the rest of the agents in social influence.

**Connectivity and Ising Model**: In the economic model, each agent is connected to all other agents, highlighting the interconnectedness of economic actors in a market in contrast to proximity effect of Ising model.

**Optimizing Surplus**: Both consumers and producers need to optimize surplus, which can be seen as maximizing benefits or profits economically. At nonzero temperatures, maximizing surplus[1] is equivalent to maximizing entropy when the temperature is fixed, given as:

$$\text{Max } \Delta(TS) = \text{Max } (S\Delta T + T\Delta S) = \text{Max } (-\Delta F + T\Delta S) \quad (4)$$

where $F$ is free energy.

**Government's Role**: The government is likened to a small group capable of raising the "temperature" (economic activity), which should benefit the entire system. This association resembles the government providing resources or stimuli to boost economic activity. Politicians are cautioned against meeting their own needs through economic activities to maintain fairness and avoid issues like bias, corruption, or bribery.

**Ferromagnetic as Social Influence:** This work suggests the behavior of ferromagnetic materials (which align their magnetic moments in the same direction) can be metaphorically understood as social influence, where individuals or groups align their behaviors or opinions due to external factors or interactions.

**3 Model**

The Hamiltonian of the system is written as:

$$H = -J \sum_{ij} S_i^z S_j^z - h_z \sum_i^N S_i^z \tag{5}$$

The positivity of J indicates ferromagnetic, whereas the negative value represents anti-ferromagnetism. $h_z$ is the external magnetic field and is greater than zero; to lower the energy, a higher energy state (down) is aligned with the magnetic field for a cold start ground state. Each site $i$ is connected to all other $N$-$1$ sites. In the following, $I$ is a 2 by 2 identity matrix.

$$S_i^Z = I \otimes I \ldots \otimes \sigma_i^Z \otimes \ldots \otimes I \tag{6}$$

The polarity of spins or energy gap of spin is protected by the Pauli $z$ matrix, in which spin down has higher energy than spin up in the presence of a magnetic field:

$$\sigma_z = \begin{bmatrix} 1 & 0 \\ 0 & -1 \end{bmatrix}. \tag{7}$$

## 4 Results and Discussions

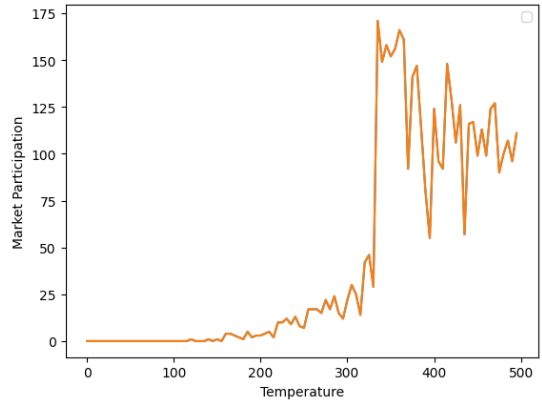

**Figure 1** For a total of 200 agents, a graphical trend of the number of buyers as a function of the temperature of the market. 80000 Monte Carlo equilibrium steps are performed at each temperature increment of 10.

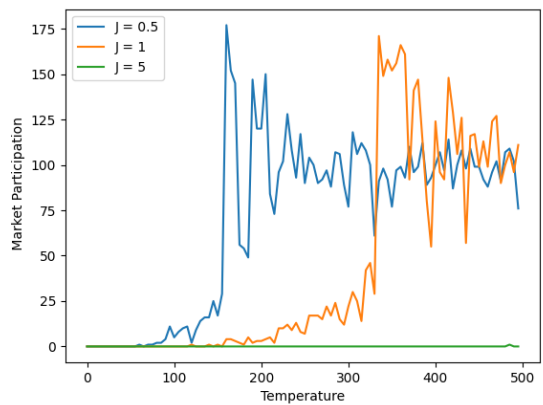

**Figure 2** For a total of 200 agents, a graphical trend of the number of buyers who have the power to participate in the market with the strength of ferromagnetic coupling represents social influences. 80000 Monte Carlo equilibrium steps are performed at each temperature increment of 10.

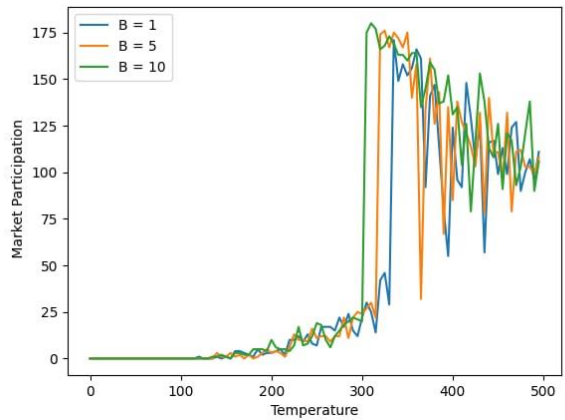

**Figure 3** For 200 agents, a graphical trend between a graphical trend of the number of buyers versus increasing temperature. The B magnetic field represents the strength of good policies to excite the market. 80000 Monte Carlo equilibrium steps are performed at each temperature increment of 10.

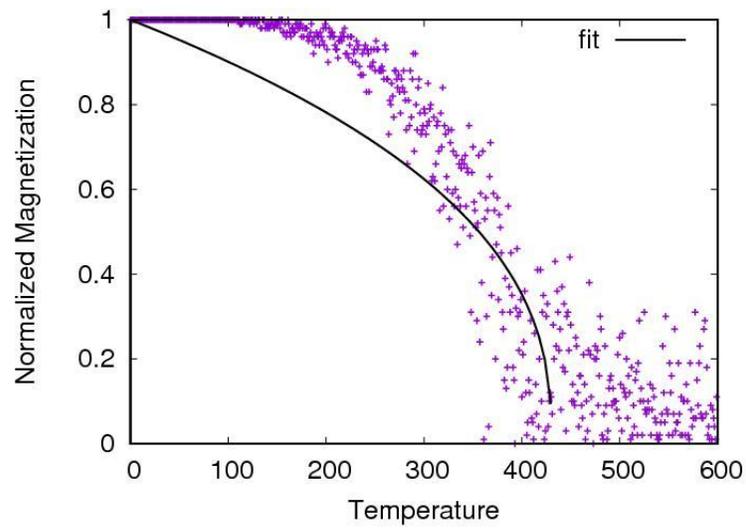

**Figure 4** For 200 agents, the magnetization vs temperature graph. 100000 Monte Carlo equilibrium steps are performed at each temperature increment of 1. Magnetization is described by Eq. (3).

The behaviour of market participants has been observed to be like the physical phenomena of magnetism in economic systems. It provides a rich analogy by considering thermodynamic principles to understand and optimize the market dynamics. This approach serves to identify how economic agents interact, and how they will make decisions and ultimately drive market outcomes. Economic agents are modelled as magnetic spins in the Ising model which means they can either engage in the activity (buy) or remain inactive (do nothing). Neighbour orientations and the coupling constant $J$ influence the spins in the Ising model. It shows an analogous to how economic agents are influenced by market conditions, individual preferences, and policies. Agents tend to minimize their activities due to constraints or uncertainty. This represents a low economic activity at low temperatures. Economic activity increased with the increase in temperature, and at this time, the agents became more engaged in buying to optimize their economic surplus.

The thermodynamics principle can be used in the sense of maximizing entropy, and the same in economics indicates the maximum surplus. If the economic system evolves freely by raising the "temperature" (economic activity), the system will approach an equilibrium, and the surplus will be optimized. By using this concept, the market dynamics can naturally evolve towards an efficient state. As in a thermodynamic system, increasing the temperature can alter spin alignments and market dynamics can be influenced significantly by government policies. Effective interventions aim to provide benefits to the entire system. Rather than serving narrow interests, these interventions will promote economic growth and stability. A powerful tool has been provided by Monte Carlo methods to simulate how market optimization evolves. The transition from low-activity to high-activity states can now be studied by modelling the economic agent as spin, and for varying levels of economic activity, their interactions can be simulated. This approach helps understand how different policies and market conditions affect the overall surplus and efficiency of the market.

Strong ferromagnetism of the model is evident as there is no sharp phase transition temperature bounded below 0.1, in the temperature regime 430 to 600 in Fig. (4). It is crucial to achieve the ideal balance without causing instability in economic systems where a significant portion of agents is active. Extreme states refer to all agents who are either inactive or fully engaged and are referred to as suboptimal. Government roles have been made to ensure interventions promote fair competition, transparency, and economic stability, creating an environment where market participants can thrive and contribute to overall growth. The limitation of the Ising model is in capturing the complex interactions between economic agents. Economic agents in this case can therefore operate as spins in a magnetic material, making it essential to accurately represent real-world economic systems to expand the model for interactions. The alignment of magnetic material due to external effects provides a metaphor for social effects in economics. The individuals or group in an economy can align their behaviours as magnetic moments align. Understanding these dynamics can help in designing economic policies.

## 5 Conclusion

This work develops a bridging concept between thermodynamics and economics to provide insight into market dynamics and optimization using mathematical models and simulations that can demonstrate how economic systems evolve and can be optimized. This interdisciplinary approach presented here both enhances theoretical frameworks and provides practical considerations to create more efficient, equitable, and sustainable markets. This article introduces a mathematical approach to simulate economic behaviors such as buying in a market context, demonstrating the role of temperature (market activity) in optimizing market dynamics and achieving equilibrium states with maximum surplus. Concepts from thermodynamics (equilibrium, non-equilibrium) combined with governance and policymaking, highlight the complexity and dynamic nature of these systems, suggesting that effective governance requires maintaining harmony with external factors, addressing imbalances, and continually adapting to meet evolving needs. The interplay between magnetic properties, temperature, and external influences on a system exhibiting ferromagnetic behaviour was discussed. The role of external factors such as magnetic fields and the value of understanding variables to achieve desired outcomes, drawing parallels to economic policies and wealth management, the optimization emphasizes the transition from ordered to disordered phases for the benefit of the markets. This work concluded that both consumer surplus and trader surplus are maximized in equilibrium by optimizing consumption/trading quantities, income, and other utility-effecting factors possibly through government interventions. These statements represent the system where temperature differences affect pricing dynamics, leading to price adjustments, surpluses, and eventually reaching an equilibrium with an optimum price level under regulated conditions.

**Additional Information**

## Code Implementation

```java
public void ising() {
    double B = 1 , J = .2 , kb = 1. , T = 1;
    Integer N = 0 ;
    Random randnum = new Random(); //Seed random generator
    PrintWriter q = null;
    try {
        prop = new Properties();
```

```java
			InputStream in = new FileInputStream(propertiesPath);
			prop.load(in);
			in.close();
	}
	catch(Exception e) {
			e.printStackTrace();
	}
	J = Double.parseDouble(prop.get("Ising_J").toString());
	kb = Double.parseDouble(prop.get("Ising_k").toString());
	T = Double.parseDouble(prop.get("Ising_T").toString());
	B = Double.parseDouble(prop.get("Ising_B").toString());
	N = Integer.parseInt(prop.get("Ising_N").toString());
	String increment = prop.get("Ising_Increment").toString();
	int upStates = 0; int downStates=0;
	int perform = Integer.parseInt(prop.get("Ising_Perform").toString());
	int deltaT = Integer.parseInt(prop.get("Ising_DeltaT").toString());
	int deltaB = Integer.parseInt(prop.get("Ising_DeltaB").toString());
	int deltaJ = Integer.parseInt(prop.get("Ising_DeltaJ").toString());
	int M = Integer.parseInt(prop.get("Ising_M").toString());
	try {
			pw = new PrintWriter ( new FileOutputStream ("ising.dat"), false );
	} catch (FileNotFoundException e) {
				// TODO Auto-generated catch block
			e.printStackTrace();
	}
	double [] state = new double[N] ; double [] test = new double[N] ;
	double ES = energy(state, N, J, B ) , prob , ET ;
	double deltaE = 0.0;
	for (int i =0 ; i < N ; i ++ ) state[ i ] = -1;
	for (int p =0; p < perform; p++) {
		for ( int j =1 ; j <= M ; j ++ ) {
			for (int k=0; k < N; k++)
				test[k] = state[k] ;
			int i=(int) ( randnum.nextDouble()*( double )N)
			test[i] *= -1;
			ET = energy (test, N, J, B ) ;
			deltaE = ET - ES;
			prob = Math.exp((ES-ET)/(kb*T)) ;
			if (deltaE < 0 || deltaE == 0) {
				for (int k=0; k< N; k++)
					state[k] = test[k] ;
				ES = ET;
			} else {
				if ( prob >= randnum.nextDouble() )
				{
					for (int k=0; k< N; k++)
						state[k] = test[k] ;
					ES = ET;
				}
			}
			upStates = 0; downStates=0;
			for ( i =0 ; i < N ; i ++ ) {
				if (state[i]==-1) {
					upStates = upStates + 1;
				} else {
```

```java
                            downStates = downStates + 1;
                        }
                    }
                }
                pw.println(T + " "+ ES+ " "+upStates + " "+downStates);
                pw.flush();
                if(increment.equalsIgnoreCase("T"))
                        T = T + deltaT;
                else if (increment.equalsIgnoreCase("H"))
                        B = B + deltaB;
                else if (increment.equalsIgnoreCase("J"))
                        J = J + deltaJ;
            }
            pw.close();
    }
    public double energy (double [] S, int Ni, double Ji, double B) {
            double FirstTerm = 0, SecondTerm = 0;
            int i;
            for (i =0 ; i <Ni ; i ++ ) {
                    for (int j = 0; j< Ni; j++) {
                        if (i!=j)
                                FirstTerm += S[i]*S[j];
                    }
            }
            FirstTerm *= -Ji ;
            for ( i =0 ; i <= (Ni-1) ; i ++ )
                    SecondTerm += S[i] ;
            SecondTerm *= -B;
            return ( FirstTerm + SecondTerm ) ;
    }
```